\documentclass[prb,twocolumn,showpacs]{revtex4-1}
\usepackage{graphicx}
\usepackage{times}
\usepackage{bm}
\usepackage{amsmath, verbatim}

\def\lsim{\mathrel{\rlap{\lower4pt\hbox{\hskip1pt$\sim$}}
    \raise1pt\hbox{$<$}}}                % less than or approx. symbol
\def\gsim{\mathrel{\rlap{\lower4pt\hbox{\hskip1pt$\sim$}}
    \raise1pt\hbox{$>$}}}                % greater than or approx. symbol

\def\be{\begin{equation}}
\def\ee{\end{equation}}
\def\bea{\begin{eqnarray}}
\def\eea{\end{eqnarray}}
\def\bse{\begin{subequations}}
\def\ese{\end{subequations}}

\def\be{\begin{eqnarray}}
\def\ee{\end{eqnarray}}

\def\ga{{\ \lower-1.2pt\vbox{\hbox{\rlap{$>$}\lower5pt\vbox{\hbox{$\sim$}}}}\ }}
\def\la{{\ \lower-1.2pt\vbox{\hbox{\rlap{$<$}\lower5pt\vbox{\hbox{$\sim$}}}}\ }}

\def\beq{\begin{equation}}
\def\eeq{\end{equation}}

\def\bea{\begin{eqnarray}}
\def\eea{\end{eqnarray}}

\begin{document}

\title{Topological minigap in quasi-one-dimensional spin-orbit-coupled semiconductor Majorana wires}

\author{Sumanta Tewari$^{1}$}
\author{T. D. Stanescu$^{2}$}
\author{Jay D. Sau$^3$}
\author{S. Das Sarma$^4$}

\affiliation{$^1$Department of Physics and Astronomy, Clemson University, Clemson, SC
29634\\
$^2$Department of Physics, West Virginia University, Morgantown, WV 26506\\
$^3$ Department of Physics, Harvard University, Cambridge, MA 02138\\
$^4$Condensed Matter Theory Center and Joint Quantum Institute, Department of Physics, University of Maryland, College Park, Maryland, 20742-4111, USA}

\begin{abstract}
The excitation gap above the Majorana fermion (MF) modes at the ends of 1D topological superconducting (TS) semiconductor wires
scales with the bulk quasiparticle gap $E_{qp}$. This gap, also called minigap,  facilitates experimental detection of the pristine TS state and MFs at experimentally accessible temperatures $T \ll E_{qp}$.
 %detection of single MFs in tunneling experiments, and also demonstrating non-Abelian statistics and TQC without
%worrying about conventional low-lying fermionic states localized at order parameter defects.
Here we show that the linear scaling of minigap with $E_{qp}$ can fail in \textit{quasi}-1D wires with multiple confinement bands when the applied Zeeman field is greater than or equal to about half of the confinement-induced bandgap. TS states in such wires have an approximate chiral symmetry supporting multiple near zero energy modes \textit{at each end} leading to a minigap which can effectively vanish. We show that the problem of small minigap in such wires can be resolved by forcing the system to break the approximate chirality symmetry externally with a second Zeeman field. Although experimental signatures such as zero bias peak from the wire ends is suppressed by the second Zeeman field above a critical value, such a field is required in some important parameter regimes of quasi-1D wires to isolate the topological physics of end state MFs. We also discuss the crucial difference
of our minigap calculations from the previously reported minigap results appropriate for idealized spinless p-wave superconductors and explain
why the clustering of fermionic subgap states around the zero energy Majorana end state with increasing chemical potential seen in the latter system does not apply to the experimental TS states in spin-orbit coupled nanowires.
 %one responsible for creating the topological state.
\end{abstract}

\maketitle

%%%%%%%%%%%%%%%%%%%%%%%%%%%%%%%%%%%%%%%%%%%%%%%%%%%%%%%%%%%%%%%%%%%%%

\section{Introduction}
Ref.~[\onlinecite{Kitaev-1D}] proposed a 1D spinless $p$-wave superconductor as a platform for end-state Majorana fermions and topological quantum computation (see also Ref.~[\onlinecite{Sengupta-2001}]).
It has been shown that a dimensionally reduced 1D version of a recently proposed 2D Rashba-coupled semiconductor heterostructure \cite{Sau} can realize such a system for
experimental investigations \cite{Unpublished,Long-PRB}. The 1D system has the advantage that the minimum excitation gap above the end-state MFs, the so-called minigap, scales with the induced quasiparticle gap $E_{qp} \sim 1$ K in the nanowire. Such a large minigap (compared with $\Delta^2/\epsilon_F \sim 0.1 \mu$K in 2D TS systems \cite{Read-Green}, for example, in Sr$_2$RuO$_4$, a potential chiral p-wave topological superconductor \cite{Sumanta-Strontium}) allows the realization of the pristine TS state and MFs at experimentally accessible temperatures by applying a strong Zeeman field. It has been shown by explicit calculations \cite{Unpublished,Long-PRB} that the end-state MFs thus produced can be probed in local tunneling experiments that evince a zero-bias conductance peak. Ref.~[\onlinecite{Oreg}], which independently proposed the 1D structure, and Ref.~[\onlinecite{Roman}] proposed an ac-Josephson experiment to probe the nanowire MFs. An alternative direct way of having a large minigap in a TS system is to use a 2D hybrid semiconductor-superconductor structure \cite{Sau} with a very small Fermi energy in the semiconductor so that the 2D minigap, i.e. $\Delta^2/\epsilon_F$,  is intrinsically large.

The basic idea behind the dimensional
reduction of the 2D structure and the associated minigap is as follows \cite{Unpublished}: The 2D $(xy)$-plane Rashba-coupled semiconductor
in the TS state supports a single gapless chiral Majorana edge mode with dispersion $E(k_y)=(\Delta/k_F) k_y$ on an edge parallel to, say, the $y$-axis. From the mathematical equivalence,
$H_{BdG}^{2D}(k_y = 0)=H_{BdG}^{1D}$, where $H_{BdG}^{1D}$ is the Bogoliubov-de Gennes (BdG) Hamiltonian of a wire along the $x$-axis, it immediately follows that a 1D nanowire along $x$ must have a zero energy  MF eigen-solution localized at the end ($E(k_y=0)=0$). For minigap, note that the 2D Hamiltonian has no other eigen-solution on the edge other than the gapless Majorana mode itself. This implies that the dimensionally reduced 1D Hamiltonian should also have no other sub-gap solution other than the zero energy state at the end. Hence, the minigap in the 1D problem should really be equal to the quasiparticle gap $E_{qp}$ induced in the nanowire. Viewed another way, in going from the 2D plane to the 1D wire, as the width $L_y$ in the $y$-direction is reduced, the energies of the quantized edge modes scale as $\sim 1/L_y$, i.e. as the inverse of the confinement size of the wire in the transverse direction -- we are assuming here that the size of the wire in the third $z$-direction is much smaller than that in the $y$-direction, but what matters is simply the transverse confinement size being very small compared with the length of the wire. Since this quantity ultimately diverges in the strict 1D limit $L_y \rightarrow 0$, it implies the minigap to be equal to the quasiparticle gap $E_{qp}$. The end-state MFs protected by such a large minigap can be probed in local tunneling experiments without having to worry about other low energy states with energies comparable to experimental temperatures. The zero energy modes should manifest themselves in zero-bias conductance peaks which exist only above a critical value of the Zeeman splitting required for the topological state. These conjectures, numerically confirmed for strict 1D wires \cite{Unpublished}, have appeared with calculational details in Ref.~[\onlinecite{Long-PRB}].

In this paper we revisit the question of the minigap in 1D wires and show that
  the linear scaling of minigap with $E_{qp}$ is valid only for strict 1D wires (or wires where the transverse confinement induced band gap far exceeds all other energy scales but excluding the Fermi energy for multiband occupancy). For quasi-1D or multi-band wires \cite{Potter-Lee} where the confinement band gap is comparable to the applied Zeeman energy scale
   \cite{Lutchyn-Stanescu,SLDS} the above linear scaling can spectacularly break down. The parameter regime where the relevant confinement band gap is comparable to the Zeeman energy
is the so-called `sweet-spot' regime \cite{Lutchyn-Stanescu,SLDS}, designed to maximize the robustness of the topological state
 to inevitable spatial chemical potential fluctuations arising, for example, from unintentional random background charged impurities in the semiconductor. We show that in this regime, because of an approximate chirality symmetry \cite{Schnyder,Kitaev-Topo-Class} of the topological state, multiple
  near zero energy modes appear at the same end and the nanowire minigap effectively vanishes. The chirality symmetry in this case refers (see Sec.~III) to the existence of a unitary operator that anti-commutes with the Hamiltonian. This symmetry,
  strictly valid for quasi-1D wires only in the absence of an interband Rashba coupling, ensures multiple zero energy MF modes on a given end of the Majorana
  wire. For a finite interband Rashba coupling the chirality symmetry is no longer exact, but we find that the quantitative effects of
  experimentally realistic interband Rashba couplings are insignificant and call the resultant Hamiltonian approximately chiral symmetric.

  Because of the possibility of multiple zero and near zero modes from the same end (with their total number given by the integer $N$ in Fig.~[1]), we use here a slightly generalized definition of the minigap as explained below. If $N$ is allowed to take only values $0,1$ (i.e., if the quantum wire had no hidden approximate chirality symmetry), the minigap would be strictly defined as the minimum excitation gap above the zero energy MF state.
  %The experimental temperature scale must be sufficiently below the minigap so the MFs do not mix with the higher energy regular fermion excitations.
  This definition of the minigap, however, is inadequate when $N$ is allowed to take \textit{any} integer values in the relevant parameter space. The integer $N$ includes, with the hidden chiral symmetry of the nanowire weakly broken by the interband Rashba term, ($N$ mod $2$) exact zero energy MF states and the remaining ($N$-($N$ mod $2$)) \textit{near zero} energy states with very small energy splittings $\sim 10^{-2} E_{qp}$, see Fig.~[3]. This is because only in the presence of exact chirality symmetry an arbitrary number of  MF modes can be supported by a given end
of a Majorana wire. A finite interband Rashba coupling breaks the chirality symmetry, splits the energies of the zero modes in pairs, but does so only weakly. So in the
presence of an experimentally realistic interband Rashba, we will show below that there are no exact zero modes for states with
even $N$, but there are $N$ near zero modes. Similarly, in states with odd $N$, $N-1$ zero modes turn into near zero modes by
the interband Rashba term, while exactly one mode remains a zero energy MF mode. The remaining single MF on a given end cannot further be split
because it is protected by the particle-hole symmetry of the BdG Hamiltonian which cannot be broken in a superconductor.~\cite{Kitaev-1D} 

 The slightly generalized definition of the minigap can be understood by noting that, when, say, the $N=1$ state (Fig.~[1]) is being probed by a zero bias tunneling peak, for an unambiguous determination of the MF one must also ensure that the tunneling in
  reality is not probing the near zero energy end-modes in the adjoining $N=2$ state. This implies that the definition of the minigap be expanded to include also the energy of the near zero modes in the parameter regime \textit{nearby} to the MF. This minigap should also be raised sufficiently above the experimental temperatures to exclude unintentional contributions from near zero energy states in the relevant parameter space. As we will show, the minigap, thus defined, effectively vanishes for the semiconductor nanowire because of the hidden approximate chiral symmetry. As in the case of 2D TS systems, it
   then becomes technically very difficult (although may not be impossible \cite{Akhmerov}) to probe the physics of isolated MFs and non-Abelian statistics at experimentally
    accessible temperatures \cite{Akhmerov,Gunnar}.

    Fortunately, in 1D, we find that there is a solution to the small minigap problem which involves forcefully breaking the nanowire chirality symmetry by applying a second Zeeman field transverse to the wire. The second Zeeman field, orthogonal to the one realizing the topological state, enhances the energies of the near zero energy modes in pairs but leaves the MF state intact, before removing all $N$ states above a critical value at which the system transitions into a topologically trivial state with $N=0$ (Fig.~[5]). Thus, if a zero bias conductance peak from an end of the nanowire \cite{Long-PRB} splits with increasing the second Zeeman field, the splitting is given by the minigap created by the second Zeeman field. For the states with various $N$ (see Fig.~[1]), a zero bias conductance peak from the $N=1$ state (with no near-zero energy mode) should show no splitting with the transverse field. The zero bias peak from the $N=2$ state splits with an applied transverse Zeeman field because of the externally induced broken chirality symmetry. The zero bias peak in the $N=3$ state is expected to decrease in height with the applied field but ($N$-($N$ mod $2$)=$2$) new low energy peaks are split off following the energies of the near-zero modes with the increasing second Zeeman field. Finally, all low bias peaks should disappear for sufficiently strong second Zeeman field as the system transitions into a topologically trivial state with $N=0$ (Fig.~[5]).
    %As we show in detail, the minigap of quasi-1D wires can thus be restored to values comparable to the experimentally accessible temperatures.

\begin{figure}[tbp]
\begin{center}
\includegraphics[width=0.36\textwidth]{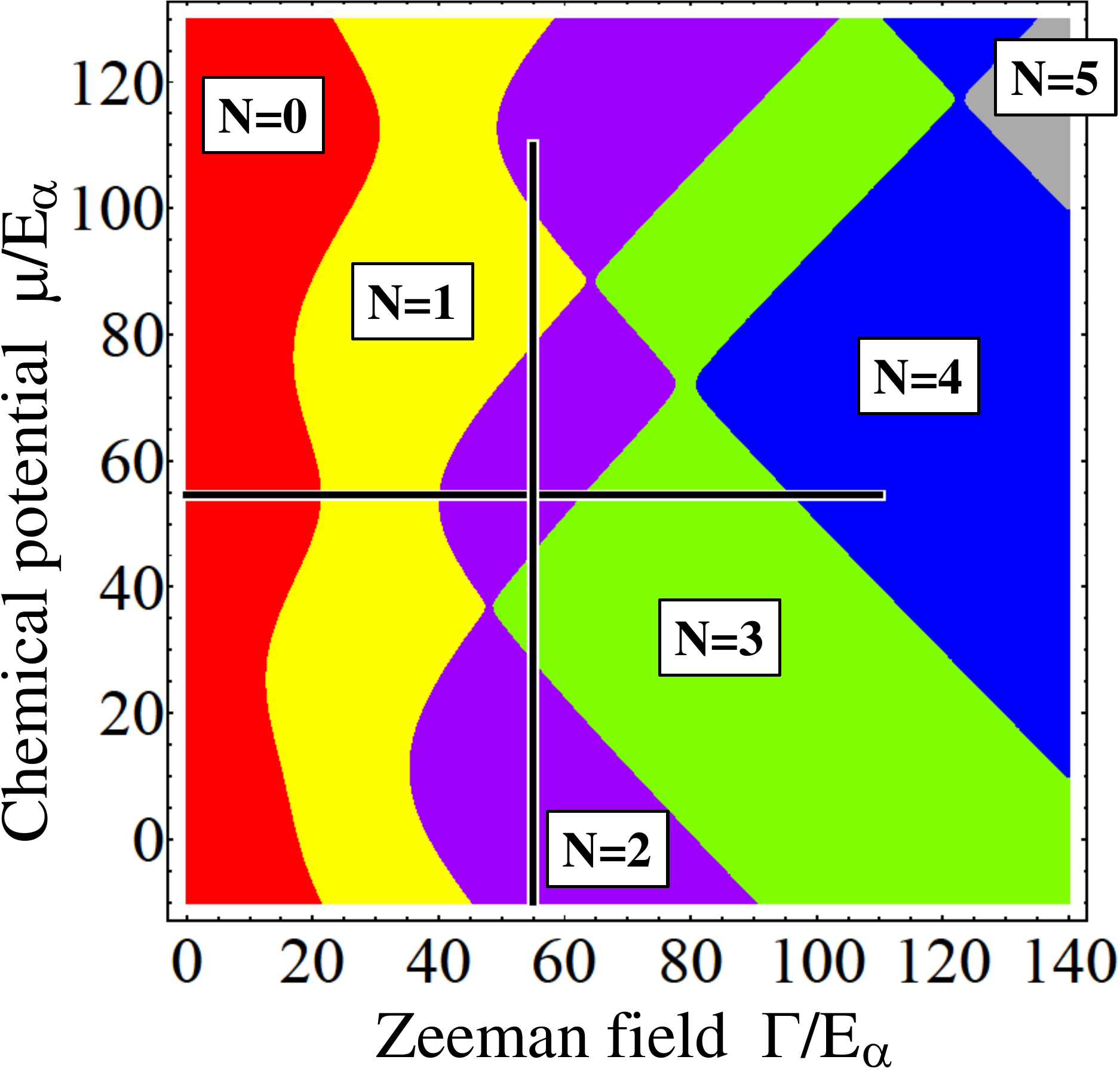}
\vspace{-5mm}
\end{center}
\caption{(Color online) Phase diagram of a quasi-1D nanowire with proximity-induced superconductivity as a function of the Zeeman field $\Gamma$ and the chemical potential $\mu$.
%The semiconductor-superconductor effective coupling is $\gamma=\Delta_0$.
%The dependence of the induced quasiparticle gap on the control parameters along the $\Gamma=const.$ and $\mu=const.$ cuts indicated by the black lines is shown in Fig. 3.
Different superconducting phases are characterized by the number $N$ of low-energy modes with $E_{\it n}\approx 0$ at each end of the wire. The energies of these modes identically vanish, with $N$ denoting the number of exact zero energy MFs at each end, for $H_{nm}(\alpha_y=0)$ (see Eq.~(\ref{Hnm})). For $H_{nm}(\alpha_y=\alpha)$,  ($N$-($N$ mod $2$)) modes are near zero modes with very small energy splittings $\sim 10^{-2}E_{qp}$ where $E_{qp}$ is the induced
quasiparticle gap $\sim 1$ K. The near-zero energy modes are very robust to all reasonable perturbations including disorder except to a strong Zeeman field perpendicular to the length of the nanowire. With the representative parameters of this paper ($E_{\alpha}= m^*\alpha^2 = 0.053$ meV and an effective $g^* \sim 50$ for InAs \cite{g-factor}), the $N=1$ region starts at $B_x \sim 0.7$ T. Due to the existence of multiple robust near zero modes at each end of the quasi-1D wire the entire region beyond $B_x \sim 0.7$ T is effectively gapless.
%For $\alpha_y\neq 0$, the number of Majorana modes is $0$ for $N$ even and $1$ for $N$ odd.
}
\vspace{-4mm}
\label{Fig2}
\end{figure}

\section{Hamiltonian, phase diagram, and multiple near zero modes in quasi-1D wires} In a recent paper \cite{Sudip} Niu \textit{et al.} have shown that a 1D
spinless $p$-wave topological superconductor can in principle support \textit{any} integer (not just $0$ or $1$) number of MFs at \textit{each end}. This has prompted two of us \cite{Sumanta-Jay-1} to examine more closely the analogous case of a 1D Rashba-coupled semiconductor nanowire with Zeeman
splitting proximity-coupled
to a s-wave superconductor.
 We have shown that this system - the so-called semiconductor ``Majorana nanowire" - can also support any integer number of MFs at a given end under invariance of the chirality symmetry.
 %This is contrary
%to the popular belief that the topological class of the semiconductor nanowire (as also of the 1D spinless $p$-wave superconductor) is D characterized
%by a $Z_2$-invariant allowing only $0$ or $1$ MFs at a single end.
 Even though the parent 2D system \cite{Sau} and others symmetry-related to it \cite{Read-Green,Zhang-Spin-Orbit,Sato-Fujimoto} are in topological class D, we have clarified that the topological class of the 1D semiconductor under chirality symmetry
is BDI. The class BDI is characterized by an integer $Z$ bulk topological invariant \cite{Schnyder,Kitaev-Topo-Class} and thus allows multiple MFs
 (with number equal to the integer invariant) even from a single end. That the topological class of the 1D system should be BDI follows from the dimensional reduction arguments mentioned in the introduction \cite{Sumanta-Jay-1}. If we think of the 1D end MF of a nanowire as the dimensionally reduced version of a 2D gapless edge MF mode, then since in 2D there are in principle $Z$ edge MF modes allowed (class D in 2D is characterized by a $Z$ invariant) it follows that there must be $Z$ end MFs allowed also in 1D. Hence the nanowire 1D system should also be with a $Z$, not $Z_2$, invariant, which leads to BDI being the appropriate topological class. (A similar dimensional reduction argument should predict that the topological class of the Fu-Kane system \cite{Fu-Kane} should reduce from DIII in 2D to D in 1D both having a $Z_2$ invariant.) To properly define the BDI class and the associated $Z$ invariant for the nanowire, we had to invoke \cite{Sumanta-Jay-1} a hidden chirality symmetry
 of the 1D system and show that, under invariance of this symmetry, the 1D nanowire can support an arbitrary integer number $N$ ($=Z$) of zero energy MFs at each end.
Below we make a departure from the case of strict 1D nanowires and consider
 the experimentally realistic case of \textit{quasi}-1D wires with Rashba and Zeeman couplings and proximity induced superconductivity. We show that it retains an \textit{approximate} chirality symmetry and thus allows the realization of multiple \textit{near zero energy} modes at each end in experimentally relevant parameter regimes.

%%%%%%%%%%%%%%%%%%%%%%%%%%%%
\begin{figure}[tbp]
\begin{center}
\includegraphics[width=0.45\textwidth]{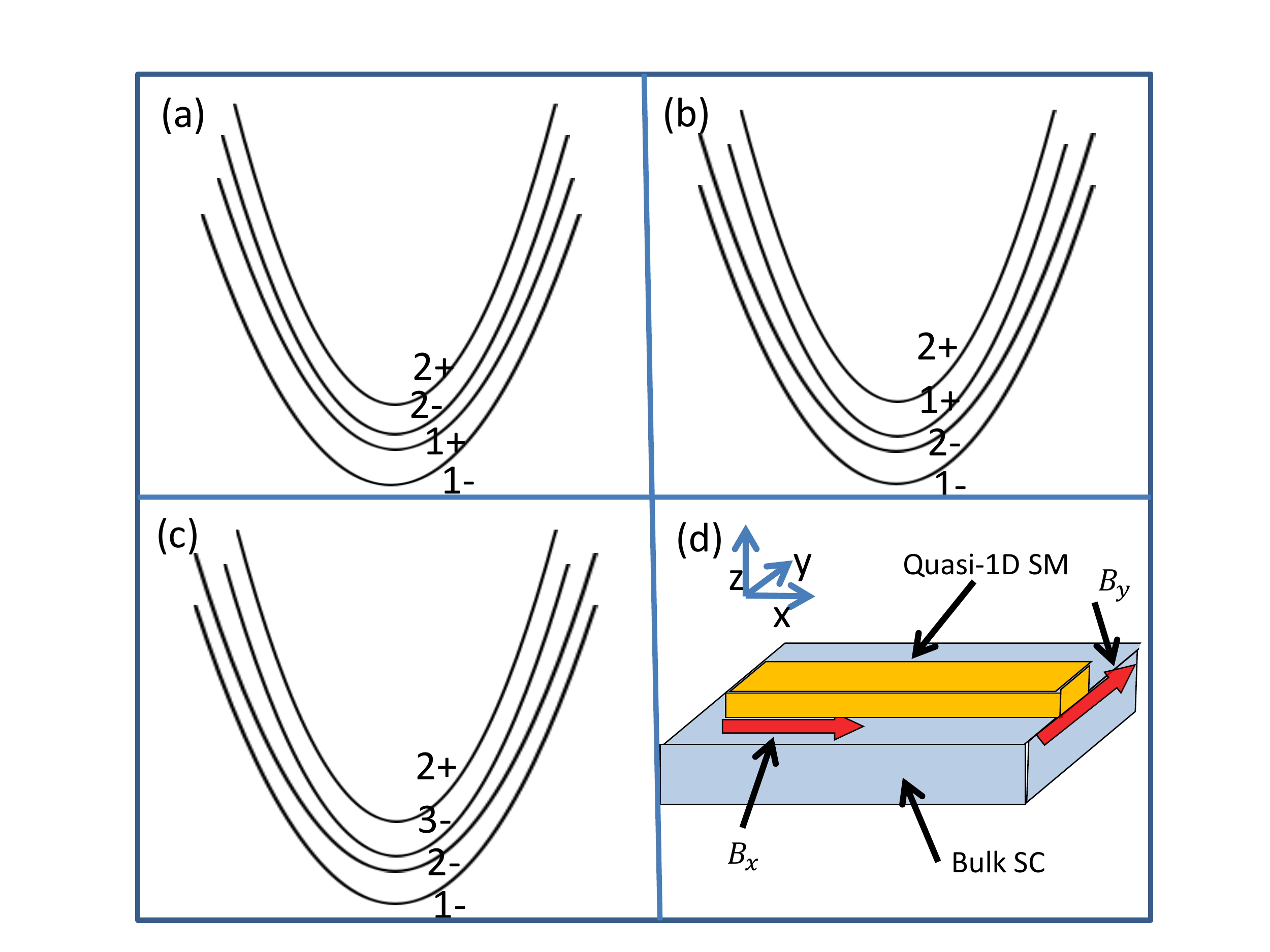}
\vspace{-5mm}
\end{center}
\caption{(Color online) Confinement band configurations for different values of the integer $N$. $N$ gives the number of near zero energy modes at each end
of quasi-1D nanowires with Hamiltonian $H_{nm}(\alpha_y=\alpha)$ given in Eq.~(\ref{Hnm}).
%$N$ indicates exact zero energy modes at each end of the nanowire with Hamiltonian $\widetilde{H}_{nm}=H_{nm}(\alpha_y=0)$. For the full Hamiltonian $H_{nm}(\alpha_y=\alpha)$, $(N {\rm{mod}} 2)$ gives the number of exact zero energy modes while the rest of the localized modes are near zero energy modes with a small energy $\sim 10^{-2}E_{qp}$.
(a) Band
configuration for $N=0,1$. $n^-$ and $n^+$ denote a pair of sub-bands for each confinement band index $n$ and move in opposite directions in response to a Zeeman field. (b) Sub-band configuration for higher Zeeman fields producing $N=2$. Note that the sub-bands $1^+$ and $2^-$ are switched in energy by the Zeeman field. (c) Confinement band configuration capable of producing $N=3$. (d) The experimental set-up consisting of a nanowire on top of a superconductor. The thick (red) arrows indicate applied magnetic fields. $B_x$ produces the TS state, while $B_y$ is an additional transverse field proposed here to externally break the chirality symmetry and produce a reasonable minigap.}
\vspace{-4mm}
\label{Fig5}
\end{figure}
%%%%%%%%%%%%%%%%%%%%%%

 We consider a rectangular semiconductor (SM)  nanowire with lengths $L_x\gg Ly\gg Lz$ proximity coupled at the interface $z=0$ to an s-wave superconductor (SC) with  a superconducting gap $\Delta_0=1$ meV.
 %We assume that the SM-SC coupling is non-homogeneous across the wire and described by the effective coupling $\gamma(y)$. In the numerical calculations we consider the intermediate coupling regime characterized by an average value of the coupling strength $\overline{\gamma}=\Delta_0$.
 The derivation of the low-energy effective Hamiltonian of the semiconductor wire that includes the induced s-wave pairing and the energy renormalizations due to proximity to the SC is described in detail in Ref.~[\onlinecite{SLDS}]. For an infinite wire (infinite in the $x$-direction), the BdG Hamiltonian of the multiband system has the form,
\begin{eqnarray}
H_{nm}(k) &=&[\epsilon_{nm}(k) -\mu \delta_{nm}]\tau_z + \Gamma\delta_{nm}\sigma_x\tau_z \nonumber \\
&+& \alpha k \delta_{nm} \sigma_y\tau_z - i \alpha_y q_{nm} \sigma_x +\Delta_{nm}\sigma_y\tau_y,      \label{Hnm}
\end{eqnarray}
where $k=k_x$ is the wave number in the $x$-direction, $\sigma_i$ and $\tau_i$ are Pauli matrices associated with the spin-1/2 and particle-hole (p-h) degrees of freedom, respectively, and we have used the basis $(u_{\uparrow}, u_{\downarrow}, v_{\uparrow}, v_{\downarrow})$ for the p-h spinors. In Eq.~(\ref{Hnm}) $n$ and  $m$ label different confinement induced sub-bands described by the transverse wave functions $\phi_n(y) \propto \sin( n\pi y/L_y)$, $\epsilon_{nm}$ describes the SM spectrum in the absence of spin-orbit coupling, $\mu$ is the chemical potential, and $\Gamma= g^* \mu_B B_x/2$ is the external Zeeman coupling along the $x$-direction which is needed to create a topological state. In the presence of an inhomogeneous SM-SC coupling, the induced superconducting pairing $\Delta_{nm}$ contains non-vanishing inter-band components. The inter-band Rashba coupling is described by the renormalized matrix elements
$q_{nm} \propto \langle \phi_n | \partial/\partial y | \phi_m \rangle$, which couple transverse states with opposite parity. Note that
in the effective Hamiltonian all energies are renormalized due to the proximity effect by a Z-factor that depends on the strength of the SC-SM coupling,
as described in Ref.~[\onlinecite{SLDS}]. The effective parameters $\epsilon_{nm}$, $q_{nm}$, $\Delta_{nm}$ are calculated numerically following the procedure described in Ref.~[\onlinecite{SLDS}]. The value of the Rashba spin-orbit coupling used in the calculations is $\alpha=0.1$ eV \AA. For the calculations of the phase diagram (see Fig.~[1]), the coefficient of the transverse Rashba coupling $\alpha_y$ that provides inter-band coupling is taken to be equal to $\alpha$ which is the experimentally relevant value. This term, however, breaks the exact chirality symmetry of $\widetilde{H}_{nm}= H_{nm}(\alpha_y=0)$ to only an approximate one for $H_{nm}$. Thus, a finite $\alpha_y$ turns the energies of some of the exact zero modes of
$\widetilde{H}_{nm}$ into only approximate zero modes for $H_{nm}$. Despite this, as we show in detail below, the energy splittings of these
 ($N$-($N$ mod $2$)) near zero modes due to $\alpha_y=\alpha$ are very small $\sim 10^{-2}E_{qp}$. On the scale of $E_{qp}$, therefore, the low-energy spectrum of $H_{nm}(\alpha_y=\alpha)$ has the zero and near zero
modes all show up as zero energy modes (see Fig.~[3]), whose number (from each end) is given by the $Z$ topological invariant calculated for $\widetilde{H}_{nm}$ in the
appropriate parameter regime. Thus, in this paper, we will calculate the integer topological invariant (the number $W$, see Eq.~(\ref{eq:W})) for $\widetilde{H}_{nm}$ and compare it with the number of near zero modes from a given end calculated by diagonalizing the full Hamiltonian $H_{nm}(\alpha_y=\alpha)$.

The Hamiltonian $H_{nm}(\alpha_y=\alpha)$ has been recently studied extensively in connection with realizing MFs in quasi-1D nanowires \cite{Lutchyn-Stanescu,SLDS}. Since the Hamiltonian explicitly takes into account multiple confinement band occupancy, it not only lifts the stringent condition of strict one-dimensionally of the nanowire but also allows higher carrier density in the topological states. Even more importantly, for values of the Zeeman coupling about half the confinement induced band gap at $k=0$, the system becomes topological over a wide range of the chemical potential $\mu$, see Fig.~[1].
%%%%%%%%%%%%%%%%%%%%%%%%%%%%
The quasi-1D nanowire, thus, allows the widest possible variation of $\mu$ without crossing a topological phase transition, leading to the maximum robustness of the end-state MFs to spatial disorder. This can be most easily understood as follows: Each confinement induced band index $n$ corresponds to
a pair of spin-orbit bands which are degenerate at $k=0$ in the absence of $\Gamma$. A finite $\Gamma$ lifts this degeneracy and the two sub-bands in a given pair,
say $\{n^{+}, n^{-}\}$, move up and down (in opposite directions along energy axis) in response to increasing $\Gamma$. When the chemical potential falls in the gap created
between $n^+$ and $n^-$ at $k=0$, the system is topological and hosts a single zero energy MF at each end of the wire. However, with increasing values of $\Gamma \sim |E_{n}(k=0,\Gamma=0)-E_{n+1}(k=0,\Gamma=0)|/2$ the energies of the bands $n^{+}$ and $(n+1)^-$ come close, and it can be shown that \cite{SLDS,Lutchyn-Stanescu} in this regime (the so-called `sweet spot' regime) the system is topological over a much wider
regime of $\mu$ than is allowed for smaller values of $\Gamma$. The wide variation of $\mu$ allowed in the same topological state can be seen, for instance, from the $N=1$ region in Fig.~[1], $N$ indicating the number of zero energy modes at each end.
\begin{comment}
It must be kept in mind that, for the full Hamiltonian $H_{nm}(\alpha_y=\alpha)$, ($N$ mod $2$) indicates the number of exact zero modes at each end, the rest being near-zero modes with energy $\sim 10^{-2}E_{qp}$. For the reduced Hamiltonian $\widetilde{H}_{nm}=H_{nm}(\alpha_y=0)$, $N$ indicates the number of exact zero energy MFs at each end.
\end{comment}

From Fig.~[1] it is clear that, with increasing values of $\Gamma$ at fixed $\mu$ (the solid horizontal line), the system makes a series of topological transitions at which the integer $N$ increases by unity. For a fixed $\Gamma$, a similar trend of discontinuous shifts of $N$ is also visible with increasing values of $\mu$ (the solid vertical line in Fig.~[1]). Later we will explain the jumps of the integer $N$ in terms of discontinuous shifts of the $Z$ topological invariant of the chiral symmetric Hamiltonian $\widetilde{H}_{nm}$. In the limit of vanishing proximity induced pair potential, the integer enhancements of $N$ with increasing $\Gamma$ can be understood as follows: For values of $\Gamma$ at which the series of bands $\{n^-,n+\}$ are `normal-ordered' (see Fig.~[2a]), the only possibilities for $N$ (equal to the integer topological invariant $W$ (eq.~(\ref{eq:W})) of $\widetilde{H}_{nm}$) are the integers $0$ and $1$. For $\Gamma >|E_{1}(k=0,\Gamma=0)-E_{2}(k=0,\Gamma=0)|/2$, the bands $1^+$ and $2^-$ switch in the energy axis, see Fig.~[2b]. In this configuration of the bands, for values of $\mu$ corresponding to two Fermi surfaces from the lowest two bands ($1^-, 2^-$), the topological $W$ number, hence $N$, becomes equal to $2$. The system thus can support two independent exact zero energy MFs at each end for $\alpha_y=0$. As mentioned before, a finite $\alpha_y=\alpha$ raises the energies of this pair of end states by a minute amount $\sim 10^{-2}E_{qp}$, where $E_{qp}$ gives the energy of the next higher excited state in the nanowire. For the full Hamiltonian $H_{nm}(\alpha_y=\alpha)$, $N=2$ therefore indicates two near zero energy states localized at each end. Proceeding this way, with further increase of $\Gamma$, when the configuration of the subbands become similar to that in Fig.~[2c] $N$ becomes equal to $3$. While $N=3$ indicates three exact MFs from each end for $\widetilde{H}_{nm}$, for $H_{nm}(\alpha_y=\alpha)$ it indicates only one end-state MF and two other near zero energy localized states with energies $\sim 10^{-2} E_{qp}$. The appearance of multiple near zero energy states at each end of the nanowire effectively reduces the minigap to $\sim 10^{-2} E_{qp}$. It is in this sense we claim that the linear scaling of the minigap with $E_{qp}$, valid in the case of strict 1D wires \cite{Unpublished,Long-PRB} or when the confinement band gap far exceeds all energy scales excluding the Fermi energy, breaks down in the case of quasi-1D wires in the experimentally relevant sweet spot regimes. We will now explain the occurrence of multiple near zero modes at
each end of the multiband wire in terms of an approximate chirality symmetry and a topological integer invariant hidden in Eq.~(\ref{Hnm}).

%%%%%%%%%%%%%%%%%%%%%%%%%%%%
\begin{figure}[tbp]
\begin{center}
\includegraphics[width=0.38\textwidth]{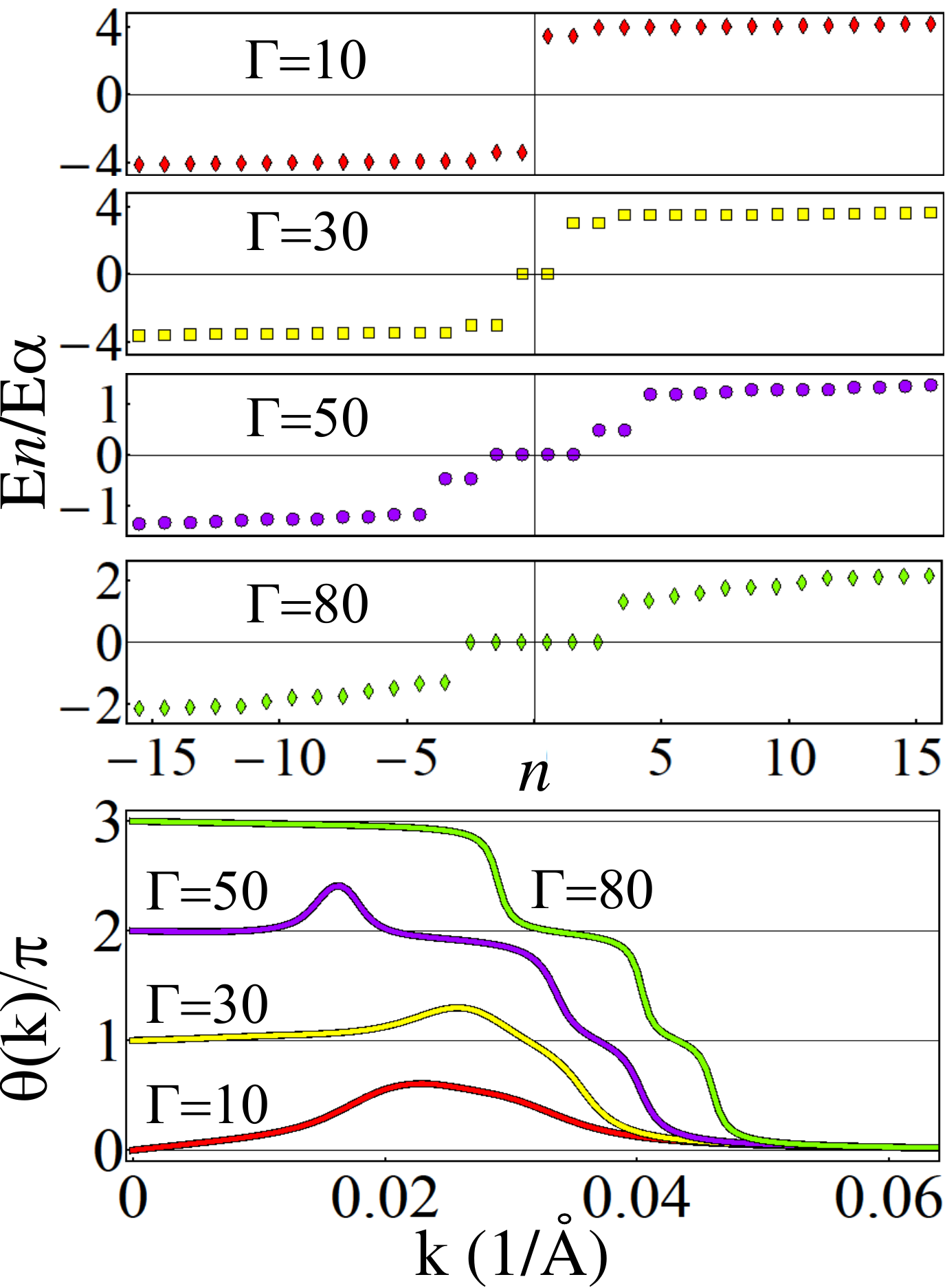}
\vspace{-5mm}
\end{center}
\caption{(Color online) Correspondence between the number of near-zero modes of $H_{nm}(\alpha_y=\alpha)$ and the $Z$ topological invariant of the chiral
 symmetric Hamiltonian $\widetilde{H}_{nm}$ for a system with $\mu = 54.5 E_\alpha$ and different values of the Zeeman field $\Gamma$. {\it Upper panel}: Low-energy BdG spectra. The number of near zero energy modes at each end of the wire increases from $N=0$ (at $\Gamma=10E_\alpha$) to $N=3$ ($\Gamma=80E_\alpha$). In addition, for certain values of the Zeeman field, there are localized in-gap states with $\Gamma$-dependent energies of the order of the bulk quasiparticle gap $E_{qp}$.
%The energies of the near-zero modes for $N=2,3$ etc. are strictly zero (MF modes) only for $\alpha_y=0$. However, even for $\alpha_y=\alpha$ they remain very close to zero energy, $\sim 10^{-2}E_{qp}$.
{\it Lower panel}: Dependence of $\theta(k)$ (Eq.~(\ref{eq:Z})) on  the wave vector $k$. The angle $\theta(k)$ is defined as a continuous function of $k$ with $\theta(\pi)=0$. Note that for $k$ much larger than the Fermi wave vector $\theta(k)$ is practically vanishes. Since $\theta(0)-\theta(k)$ is an antisymmetric function of $k$ on the interval $-\pi\leq k\leq \pi$, the winding number $W$ is given by the angle at $k=0$, $W = \theta(0)/\pi$.
}
\vspace{-4mm}
\label{Fig1}
\end{figure}
%%%%%%%%%%%%%%%%%%%%%%

\begin{figure}[tbp]
\begin{center}
\includegraphics[width=0.38\textwidth]{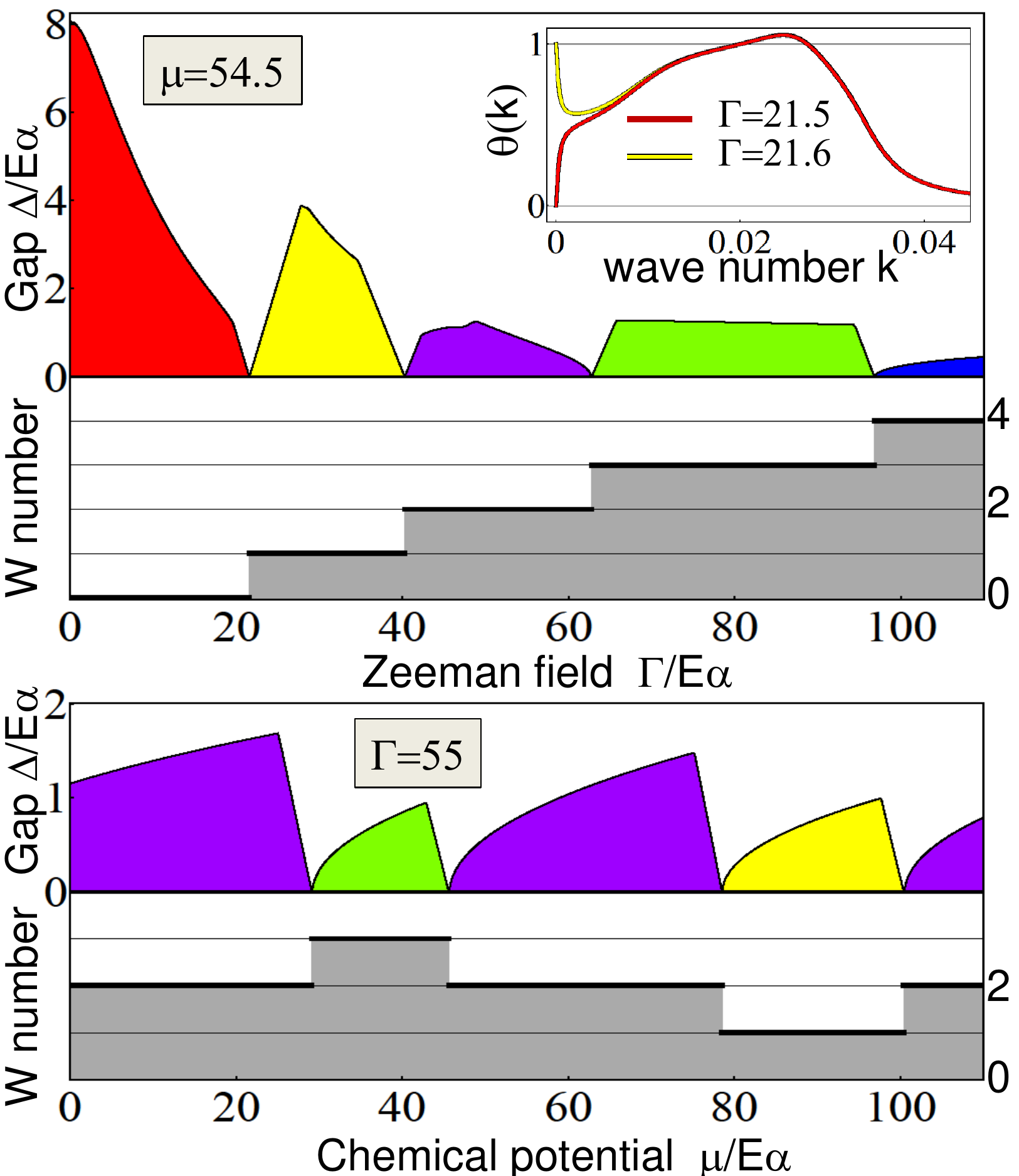}
\vspace{-5mm}
\end{center}
\caption{(Color online) Dependence of the quasiparticle gap and the integer topological invariant $W$ on the Zeeman field (upper panel) and the chemical potential (lower panel).  The parameters correspond to the horizontal and vertical cuts through the phase diagram indicated in Fig. [1]. The winding number $W$ is calculated in the absence of transverse Rashba coupling ($\alpha_y=0$), while the quasiparticle gap is determined using the full Hamiltonian $H_{nm}(\alpha_y=\alpha)$ in Eq.~(\ref{Hnm}).  Note that the phase boundaries can be identified by either the vanishing of the gap or the discontinuities of the winding number. The inset (upper panel) shows the behavior of $\theta(k)$ near the topological phase transition at  $\mu=54.5E_\alpha$ and $\Gamma\approx 21.55E\alpha$. Note that along the phase boundaries $\theta(k)$ becomes ill-defined at $k=0$.
}
\vspace{-4mm}
\label{Fig3}
\end{figure}

\section{Chirality symmetry and $Z$ invariant description of the phase diagram}
For a discussion of the chirality symmetry of the nanowire we first take $\alpha_y=0$. We will show below that this transverse Rashba term that couples the
confinement induced transverse bands indexed by $n,m$ breaks the chirality symmetry and produces a tiny minigap $\sim 10^{-2}E_{qp}$.
\begin{comment}
To enhance the minigap
 up to the experimental energy scales, later we will also include a transverse Zeeman coupling $\Gamma_y$. We will show that even though $\Gamma_y
 = g^{*}\mu_B B_y$ where $B_y$ is a transverse magnetic field is not necessary to produce the topological state and MFs, such an additional Zeeman coupling
 also breaks the chirality symmetry and enhances the minigap by an order of magnitude to make the MFs accessible in experiments.
 \end{comment}
  We consider the Hamiltonian $\widetilde{H}_{nm}=H_{nm}(\alpha_y=0)$. It can be seen by explicit construction that $\widetilde{H}_{nm}$ anticommutes with a unitary operator
 ${\cal{S}}=\tau_x$,
 \begin{equation}
 \{\widetilde{H}_{nm},{\cal{S}}\}=0.
 \label{Chirality}
 \end{equation}
 Here, the `chirality' symmetry operator ${\cal{S}}=\tau_x$ can be written as the product of an artificial `time reversal' operator ${\cal{K}}$
 and a particle-hole transformation operator $\Lambda=\tau_x\cdot{\cal{K}}$ where ${\cal{K}}$ is just the complex conjugation operator. It is easy to check explicitly that $\widetilde{H}_{nm}$ commutes with the complex conjugation operator ${\cal{K}}$ and anticommutes with the p-h transformation operator $\Lambda$, and hence it anticommutes
 with the chirality operator ${\cal{S}}={\cal{K}}\cdot\Lambda=\tau_x$. The existence of all three symmetries - `time reversal', particle-hole, and chirality - ensures that $\widetilde{H}_{nm}$ is in the BDI symmetry class \cite{Schnyder,Kitaev-Topo-Class} characterized by an integer topological invariant which we call $W$.  From Eq.~(\ref{Chirality}) it follows that the large square matrix Hamiltonian $\widetilde{H}_{nm}$ can be off-diagonalized in a basis in which the unitary operator $\cal{S}$ is diagonal:
 \begin{equation}
U \widetilde{H}_{nm}(k) U^\dagger = \left(\begin{array}{cc}0&A(k)\\ A^T(-k) & 0\end{array}\right).
\label{eq:Off-Diag}
\end{equation}
Following Ref.~[\onlinecite{Sumanta-Jay-1}] we now define the variable,
\begin{equation}
z(k)=\exp(i\theta(k))=Det(A(k))/|Det(A(k))|,
\label{eq:Z}
\end{equation}
 and calculate the integer invariant,
 %Since $p(k)$ having a zero eigenvalue implies a gap closure of $H(k)$, the winding number
%of the phase of $Det(p(k))$ i.e.
\begin{equation}
W=\frac{-i}{\pi}\int_{k=0}^{k=\pi} \frac{d z(k)}{z(k)},\label{eq:W}
\end{equation}
which is an integer ($W\in Z$) including zero.

Note that in the presence of a finite transverse Rashba coupling, $H_{nm}(\alpha_y \neq 0)$ does not anti-commute with $\tau_x$. Hence, the Hamiltonian matrix is no longer off-diagonalizable in the diagonal basis of ${\cal S}$ and the number $W$ cannot even be defined. A finite $\alpha_y$ thus breaks the chirality symmetry.
However, even though the invariant $W$ cannot be defined for a finite $\alpha_y$, since the experimental value of $\alpha_y=\alpha$ makes only a minute contribution $\sim 10^{-2}E_{qp}$
to the energies of the near zero energy end states (compared to the energies $\sim E_{qp}$ of the next higher order excitations), we calculate the $Z$ invariant for $\widetilde{H}_{nm}$ (i.e., by taking $\alpha_y=0$) and compare it with the phase diagram derived
by computing the low energy spectrum of the full Hamiltonian $H_{nm} (\alpha_y=\alpha)$ in Eq.~(\ref{Hnm}). The results are shown in Fig.~[3]. It is clear that the different topological phases of the full Hamiltonian $H_{nm} (\alpha_y=\alpha)$ characterized by different numbers $N$ of near zero energy end states can be characterized by different values of the integer $W$ calculated for the corresponding reduced Hamiltonian $H_{nm}(\alpha_y=0)$.

The accurate correspondence of the topological quantum phase transitions separating phases with different values of $N$ with the change of the integer invariant
$W$ is shown even more clearly in Fig.~[4]. In this figure, we have plotted the induced quasiparticle gap $E_{qp}$ with the tuning parameters $\Gamma$ (upper panel) and the chemical potential $\mu$ (lower panel). Phases with different values of $N$ are separated by points in the parameter space where $E_{qp}$ vanishes. As is clear from Fig.~[4] these are also the points at which the integer $W$ changes discontinuously, and the value of $W$ for a finite value of $E_{qp}$
corresponds to the number $N$ of the near zero energy modes localized at each end of the quasi-1D nanowire.

\section{Chirality breaking and non-zero minigap with transverse Zeeman coupling}
\begin{comment}
So far in our calculations of the topological invariant $W$ we have neglected
the transverse Rashba coupling $\alpha_y$. The reason for this is simple: with $\alpha_y=0$ the quasi-1D nanowire Hamiltonian in Eq.~(\ref{Hnm}) is purely real (in real space) and this ensures that the Hamiltonian commutes with the artificial `time reversal' operator ${\cal{K}}$ which leads to the existence of the chirality symmetry
operator ${\cal{S}}={\cal{K}}\Lambda$. A non-zero $\alpha_y=\alpha$ breaks the reality condition of the Hamiltonian and thus it also breaks the chirality symmetry. It is also clear from inspection of Eq.~(\ref{Hnm}) that $H_{nm}$ is no longer off-diagonalizable by the unitary operator ${\cal{S}}=\tau_x$. Therefore, the integer topological invariant $W$ in Eq.~(\ref{eq:W}) can no longer be defined for the quasi-1D wire. Nevertheless,
\end{comment}
As we have seen above, despite breaking the chirality symmetry, the quantitative effects of $\alpha_y=\alpha$ on the energies of the localized end modes
 of quasi-1D nanowires are minimal.
In particular, even though the state with $N=2$ no longer has an exact zero energy mode localized at the ends, the energy splitting caused by $\alpha_y$ is
very small, $\sim 10^{-2}E_{qp}$ ($E_{qp} \sim 1$ K). The situation is similar for $N=3$ where $\alpha_y=\alpha$ raises the energies of two of the zero modes to
$\sim 10^{-2}E_{qp}$ while the third one remains at exact zero energy which is a Majorana mode. We emphasize that the lowest energy state for $N=3$ is
a Majorana mode. Any finite gap acquired by this state is entirely due to
finite size effects. \cite{Aguado} The finite transverse Rashba coupling thus creates
a small minigap above the end state MFs, but such a small gap is difficult to resolve experimentally at experimentally realistic temperatures.

%The problem of the small minigap in quasi-1D wires near the experimentally important sweet spot regime can be resolved by forcing the system to
%break the chirality symmetry externally.

%%%%%%%%%%%%%%%%%%%%%%%%%%%%
\begin{figure}[tbp]
\begin{center}
\includegraphics[width=0.35\textwidth]{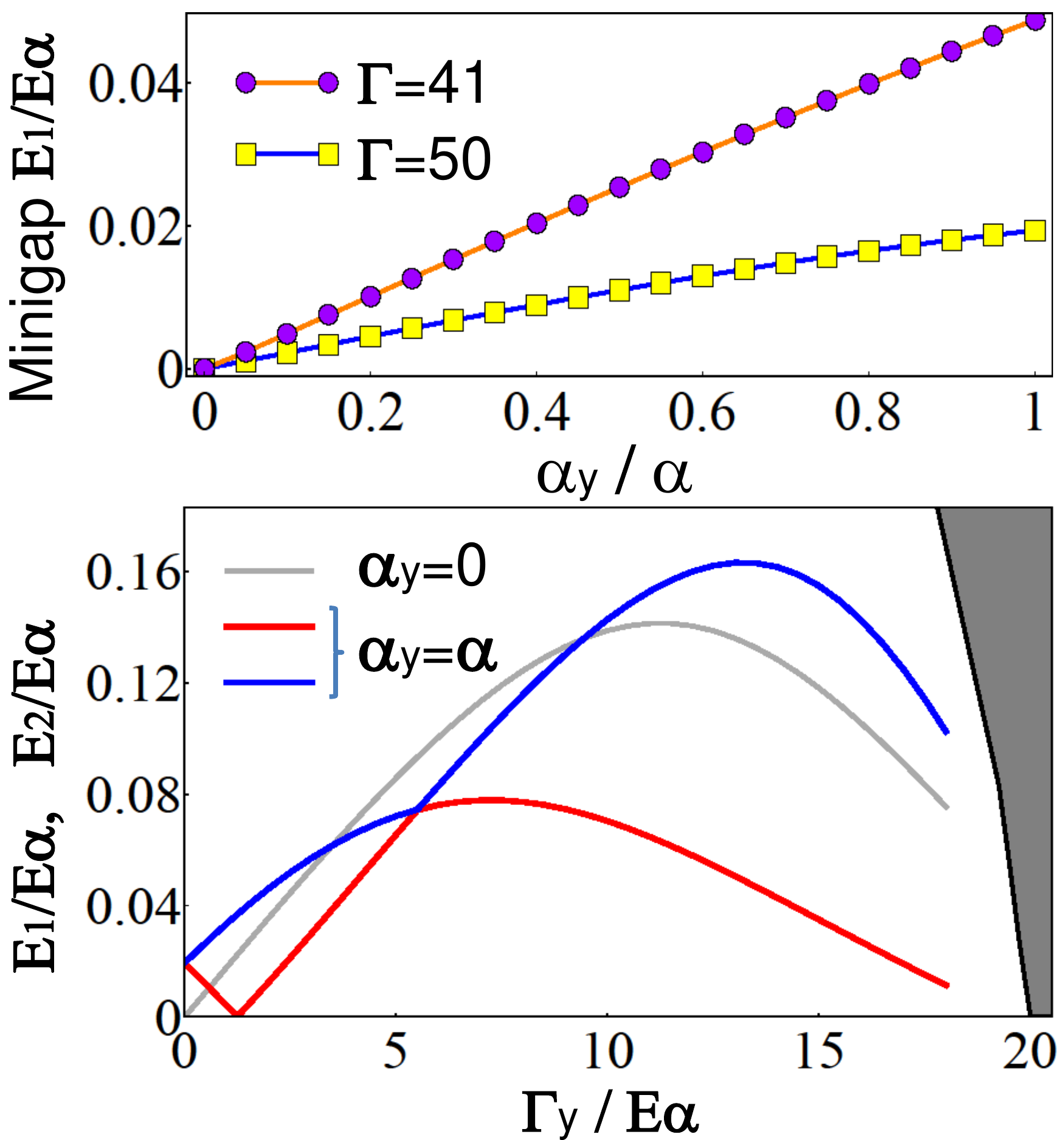}
\vspace{-5mm}
\end{center}
\caption{(Color online) {\it Upper panel}: Dependence of the minigap on the transverse Rashba coupling $\alpha_y$ for $\mu=54.5E_\alpha$ and two different  values of the Zeeman field in the phase with $N=2$. Note that the low-energy modes have a finite energy $E_1=E_2$ that increases approximately linearly with the transverse Rashba coupling strength. For $\alpha_y=\alpha$ the minigap attains values $\sim 1-2\times 10^{-2}E_{qp}$ (see Fig.~[4] for comparison).  {\it Lower panel}: Dependence of the  lowest energy states on a transverse Zeeman field for $\mu=54.5E_\alpha$ and  $\Gamma= 50 E_\alpha$. Note that $E_1=E_2$ if $\alpha_y=0$, while $E_1\neq E_2$ in the presence of both the transverse Zeeman field $\Gamma_y$ and the transverse Rashba coupling $\alpha_y$. For $\Gamma_y > 20E_\alpha$ the quasiparticle gap itself collapses (black line) and the nanowire makes a transition to a topologically trivial phase with no MFs.
}
\vspace{-4mm}
\label{Fig5}
\end{figure}
%%%%%%%%%%%%%%%%%%%%%%

We propose to solve the small minigap problem of quasi-1D nanowires by applying an additional \textit{transverse} Zeeman field $\Gamma_y=g^*\mu_BB_y/2$ in addition to the longitudinal one needed to create the TS state itself (see Fig.~[2d]). Note that such a Zeeman field is still parallel to the plane of the nearby bulk superconductor and thus does not create additional problems with the s-wave superconducting pairing. In the presence of this term, the BdG Hamiltonian of the nanowire becomes,
\begin{equation}
H^{\prime}_{nm}(k) = H_{nm}(k) + \Gamma_y\delta_{nm}\sigma_y.      \label{H'nm}
\end{equation}

It can be easily checked that both terms with coupling constants $\Gamma_y$ and $\alpha_y$ retain diagonal elements even in the basis in which
${\cal S}=\tau_x$ is diagonal. This implies that $H^{\prime}_{nm}$ is no longer chiral symmetric with ${\cal S}$ as the chirality operator. Based on the available discrete symmetries (time reversal, particle-hole, complex conjugation, etc.) it is not possible to find any unitary symmetry operator that anticommutes with $H^{\prime}_{nm}$. It follows that the chiral symmetry hidden in $\widetilde{H}_{nm}$ has now been broken \textit{externally} by $\Gamma_y$.
 In the absence of $\Gamma_y$ if the number of (near) zero modes is even,
the transverse field creates a gap for all of them resulting in no MF edge mode.
In cases where the number of zero modes is odd for $\Gamma_y=0$, the transverse field opens a gap
for all of them save one, and the system is reduced to having only one non-degenerate
MF edge mode (at each end). As a consequence, the minigap above the end-state MFs is now externally tunable by tuning $\Gamma_y$. It should be kept in mind, however, that $\Gamma_y$ cannot be increased to arbitrarily large values to attain a large minigap $\sim E_{qp}$. This is because, above a critical value $\Gamma_y=\Gamma_y^c$, the quasiparticle gap $E_{qp}$ itself closes and the system makes a transition into a topologically trivial state with no MFs (see Fig.~[5]). Keeping these constraints in mind, we show in Fig.~[5] that a reasonable $\Gamma_y$
($<\Gamma_y^c$) can raise the minigap to about $\sim 0.1 E_{qp}\sim 100$ mK.

For experimental signatures such as the zero bias conductance peak from wire ends \cite{Long-PRB}, our
analysis predicts that in the sweet spot regimes for $\Gamma_y=0$ the zero bias peak in the $N=1$ state of Fig.~[1] is expected to continue also in the
$N=2$ state because of the existence of two near zero modes from the same end. In the presence of increasing $\Gamma_y$, the peak in the $N=1$ state will
disappear above $\Gamma_y > \Gamma_y^c$. The zero bias peak in the $N=2$ state, however, is expected to first split into two with increasing $\Gamma_y$ (
because of the splitting (Fig.~[5]) induced by the broken chirality symmetry), before disappearing altogether for $\Gamma_y > \Gamma_y^c$ in the topologically
trivial state.
\begin{comment}
 $\Gamma_y$ is expected to monotonically \textit{reduce} the zero bias peak, before removing it altogether for
$\Gamma_y > \Gamma_y^c$. This is because a finite $\Gamma_y$ removes ($N$-($N$ mod $2$)) chiral symmetric near zero modes from a single end, which are otherwise robust to all
reasonable perturbations including disorder, before removing the ($N$ mod $2$) MFs themselves in the topologically trivial phase above $\Gamma_y^c$.
 \end{comment}
 Thus, just for the observation of the zero bias peak from the wire ends, a non-zero $\Gamma_y$ is not required. However, to unambiguously probe the physics of isolated MFs and non-Abelian statistics in the sweet spot regimes, an optimum transverse Zeeman field is required to raise the minigap sufficiently above the experimentally accessible temperatures.

\section{Conclusion} The minigap above the end-state Majorana fermions in 1D topological superconducting nanowires scales linearly with the bulk quasiparticle gap $E_{qp}$. This fact, now confirmed in many numerical calculations \cite{Unpublished,Long-PRB,SLDS,Chuanwei-Hole-Doped}, solves the problem
of small minigap ($\sim \Delta^2/\epsilon_F \sim 0.1 \mu$K) of 2D TS systems where MFs appear in bulk order parameter defects such as vortices. Here we show that the linear scaling of minigap with $E_{qp}$ is valid only for strict 1D wires or where the confinement induced band gap far exceeds all other
energy scales excluding the Fermi energy. We show that the linear scaling fails in parameter regimes (called `sweet spot' regime) where
the confinement band gap is in the range of the Zeeman energy scale for which the MFs have been proposed \cite{Lutchyn-Stanescu,SLDS} to be most stable against spatial disorder and chemical potential fluctuations. The failure of the linear scaling arises from an approximate chiral symmetry of quasi-1D nanowires leading to the possibility
of multiple near zero energy modes from each end. The near zero modes are robust to all reasonable perturbations including disorder, except to an additional transverse Zeeman field that can be used to break the approximate chiral symmetry externally. We show that the nearly vanishing minigap of quasi-1D nanowires in this regime can be restored to experimentally accessible temperatures by applying the second Zeeman field perpendicular to the length of the nanowire. For experimental signatures such as the zero bias conductance peak from the wire ends \cite{Long-PRB}, the additional Zeeman field removes the zero bias anomaly in the topologically trivial phase for $\Gamma_y > \Gamma_y^c$. An optimum $\Gamma_y$ is needed, however, to probe the topological physics of isolated MFs in sweet spot regimes because it can raise the minigap above the experimentally
realistic temperatures.

The present paper provides the theoretical description of the minigap for the experimentally relevant quasi-1D semiconductor quantum wires with strong spin-orbit coupling in proximity to a superconductor and external Zeeman fields. Previously reported calculations for the minigap \cite{Brouwer,Potter-Minigap} based on idealized quasi-1D spinless p-wave superconducting wires
do not apply to the experimental semiconductor nanowire systems where the interplay of spin-orbit coupling and Zeeman splitting leads to a helical topological superconductor in the presence of the proximity effect. Even though the two systems have some superficial similarities, notably the topological class which is BDI under
chirality symmetry \cite{Sumanta-Jay-1}, the calculation of the minigap is one example where the difference is explicit as explained below. For a quasi-1D spinless p-wave superconductor Ref.~[\onlinecite{Brouwer}] finds a large number of end-localized subgap states with very low energies clustering around the zero energy end state when the chemical potential crosses many confinement induced subbands. The interpretation of this effect \cite{Brouwer} is in terms of each subband contributing its own MF state, which are
then weakly split by the transverse couplings leaving only one exact zero energy eigenvalue when the number of relevant subbands is odd. We want to emphasize that this effect, which drastically reduces the minigap
in spinless p-wave wires with many occupied subbands, is completely absent in the spin-orbit coupled semiconductor wires at low applied fields. For the latter system it is clear from Fig.~[3] (upper panel,
$\Gamma=30 E_{\alpha}$) that despite many subband crossings the energy of the first excited state above the MF end state for $N=1$ is close to the
bulk quasiparticle gap $E_{qp}$. Thus, in this case, there is no effect akin to the crowding near zero energy effect discussed in the context of spinless p-wave superconductors. Therefore, the fermionic near-zero energy states of Ref.~[\onlinecite{Brouwer}] are not a problem for the topological phase in real semiconductor nanowires although they are generically present in the Kitaev-type idealized model of spinless 1D p-wave superconductor.  Thus, the experimental semiconductor nanowires are in fact better systems for realizing Majorana modes than the idealized spinless p-wave 1D system often studied in theoretical analysis.

This crucial difference between the spinless p-wave case and the case of a semiconductor quantum wire with spin-orbit coupling can be understood with the help of the schematic energy level diagrams in Fig.~[2]. Note that the effective p-wave superconducting order parameter in the basis of the semiconductor bands
comes with a sign difference between the two bands in a given spin-orbit pair \{$n^{-}, n^{+}$\}. This effect underlies the fact that with increasing $\mu$,
but staying at a fixed $\Gamma$, no extra near zero modes are generated in the wire ends despite more and more subbands being populated with increasing chemical potential. This is no longer true if the Zeeman coupling is large enough to switch the order of the subbands (see Fig.~[2]) and only bands with one sign of the effective superconducting order parameter are populated with increasing $\mu$. In this case, additional MFs will appear with increasing number of the occupied subbands, which will then be weakly split by the chirality breaking fields we discuss in the paper. While this is the case with the spin-orbit coupled quantum wires, the quasi-1D spinless p-wave superconductor
 has the same sign of the order parameter in all its confinement bands. This places the ideal 1D p-wave system considered in Ref.~[\onlinecite{Brouwer}] in the $\Gamma=\infty$ limit
 of the spin-orbit coupled quantum wires and only in this unphysical limit the minigap is reduced in the experimental quantum wires by the multiband effects discussed in the
  context of spinless p-wave superconductors. This problem obviously will not arise in the $N=1$ state (see Fig.~[1]) of the real experimental systems which are spin-orbit coupled semiconductor nanowires.

This work is supported by DARPA-MTO, NSF, DARPA-QuEST, JQI-NSF-PFC, Harvard Quantum Optics Center, and Microsoft-Q.


\begin{thebibliography}
%{Bangura et~al.(2008)Bangura, Fletcher, Carrington,
%Levallois, Nardone, Vignolle, Heard, Doiron-Leyraud, LeBoeuf, Taillefer
%et~al.}
\bibitem{} \expandafter\ifx\csname natexlab\endcsname\relax

\fi
\expandafter\ifx\csname bibnamefont\endcsname\relax

\fi
\expandafter\ifx\csname bibfnamefont\endcsname\relax

\fi
\expandafter\ifx\csname citenamefont\endcsname\relax

\fi
\expandafter\ifx\csname url\endcsname\relax

\fi
\expandafter\ifx\csname urlprefix\endcsname\relax

\fi
\providecommand{\bibinfo}[2]{#2} \providecommand{\eprint}[2][]{\url{#2}}

\bibitem{Kitaev-1D} A. Y. Kitaev, Physics-Uspekhi \textbf{44}, 131 (2001).

\bibitem{Sengupta-2001} K. Sengupta, I. Zutic, H.-J. Kwon, V. M. Yakovenko, S. Das Sarma, Phys. Rev. B \textbf{63}, 144531 (2001).

\bibitem{Sau} Jay D. Sau, R. M. Lutchyn, S. Tewari, S. Das Sarma,
 Phys. Rev. Lett. \textbf{104}, 040502 (2010).

 \bibitem{Unpublished} S. Tewari, J. D. Sau, S. Das Sarma, unpublished (2009).

 \bibitem{Long-PRB} J. D. Sau, S. Tewari, R. Lutchyn, T. Stanescu and S. Das Sarma, Phys. Rev. B \textbf{82}, 214509 (2010).

 \bibitem{Read-Green} N. Read and D. Green, Phys. Rev. B \textbf{61}, 10267
(2000).

\bibitem{Sumanta-Strontium} S. Das Sarma, C. Nayak, S. Tewari, Phys. Rev. B \textbf{73}, 220502 (R) (2006).

 \bibitem{Oreg} Y. Oreg, G. Refael, F. V. Oppen, Phys. Rev. Lett. \textbf{105}, 177002 (2010).

  \bibitem{Roman} R. M. Lutchyn, Jay D. Sau, S. Das Sarma, Phys. Rev. Lett. \textbf{105}, 077001 (2010).

  \bibitem{Potter-Lee} A. C. Potter, P. A. Lee, Phys. Rev. Lett. \textbf{105}, 227003 (2010).

  \bibitem{Lutchyn-Stanescu} R. M. Lutchyn, T. D. Stanescu, S. Das Sarma, Phys. Rev. Lett. \textbf{106}, 127001 (2011).

  \bibitem{SLDS} T. D. Stanescu, R. M. Lutchyn, S. Das Sarma, Phys. Rev. B \textbf{84}, 144522 (2011).


  \bibitem{Schnyder}A. P. Schnyder, S. Ryu, A. Furusaki, and A. W. W. Ludwig, Phys. Rev. B \textbf{78} 195125 (2008);
A. P. Schnyder, S. Ryu, A. Furusaki, and A. W. W. Ludwig,  AIP Conf. Proc. \textbf{1134} 10 (2009).

\bibitem{Kitaev-Topo-Class} A. Yu Kitaev AIP Conf. Proc. \textbf{1134} 22 (2009).

\bibitem{Akhmerov} A. R. Akhmerov, Phys. Rev. B \textbf{82}, 020509 (2010).

\bibitem{Gunnar} G. M\"{o}ller, N. R. Cooper, V. Gurarie, Phys. Rev. B \textbf{83}, 014513 (2011).

\bibitem{Sudip} Y. Niu, S.-B. Chung, C.-H. Hsu, I. Mandal, S. Raghu, S. Chakravarty, Phys. Rev. B \textbf{85}, 035110 (2012).

\bibitem{Sumanta-Jay-1} S. Tewari, J. D. Sau, arXiv: 1111.6592v2.


\bibitem{g-factor} S. Nadj-Perge, V. S. Pribiag, J. W. G. van den Berg, K. Zuo, S. R. Plissard, E. P. A. M. Bakkers, S. M. Frolov, L. P. Kouwenhoven,
arXiv: arXiv:1201.3707.



\bibitem{Fu-Kane} L. Fu, C. L. Kane, Phys. Rev. Lett. \textbf{100}, 096407 (2008)

\bibitem{Zhang-Spin-Orbit}C. W. Zhang, S. Tewari, R. M. Lutchyn, S. Das Sarma, Phys. Rev. Lett. \textbf{101}, 160401 (2008).

\bibitem{Sato-Fujimoto} M. Sato, Y. Takahashi, S. Fujimoto, Phys. Rev. Lett. \textbf{103}, 020401 (2009).

\bibitem{Chuanwei-Hole-Doped} L. Mao, M. Gong, E. Dumitrescu, S. Tewari, C. W. Zhang, Phys. Rev. Lett. (in press); arXiv:1105.3483v4.

\bibitem{Aguado} J. S. Lim,  L. Serra,  R. Lopez,  R. Aguado,
arXiv: 1202.5057.

\bibitem{Brouwer} G. Kells, D. Meidan, and P. W. Brouwer, Phys. Rev. B \textbf{85}, 060507(R) (2012)

\bibitem{Potter-Minigap} A. C. Potter and P. A. Lee, Phys. Rev. B \textbf{85}, 094516 (2012)


























\end{thebibliography}
\end{document}